\documentstyle[preprint,aps]{revtex}
\tighten
\draft
\widetext
\input epsf
\preprint{HUTP-99/A025, NUB 3201}
\bigskip
\bigskip
\begin{document}
\title{A Remark on Brane Stabilization in Brane World}
\medskip

\author{Zurab Kakushadze\footnote{E-mail: 
zurab@string.harvard.edu. Address after September 1, 1999: C.N. Yang
Institute for
Theoretical Physics, State University of New York, Stony Brook, NY 11794.}}

\bigskip
\address{
Jefferson Laboratory of Physics, Harvard University,
Cambridge,  MA 02138\\
and\\
Department of Physics, Northeastern University, Boston, MA 02115}
\date{June 28, 1999}
\bigskip
\medskip
\maketitle

\begin{abstract}
{}In this note we discuss dynamical mechanisms for brane stabilization
in the brane world context. In particular, we consider supersymmetry 
preserving brane stabilization, and also brane stabilization accompanied by
supersymmetry breaking. These mechanisms are realized in some four 
dimensional ${\cal N}=1$ supersymmetric orientifold models. For 
illustrative purposes we consider two explicit orientifold models previously
constructed in \cite{class}. In both of these models branes are stabilized at
a finite distance from the orientifold planes. In the first model brane 
stabilization occurs via supersymmetry preserving non-perturbative gauge
dynamics. In the second model supersymmetry is dynamically broken, and 
brane stabilization is due to an interplay between non-perturbatively generated
superpotential and tree-level K{\"a}hler potential.
   
\end{abstract}

{}The discovery of D-branes \cite{polchi} is likely to have important 
phenomenological implications. Thus, the Standard Model gauge and matter 
fields may live inside of $p\leq 9$ spatial dimensional $p$-branes, 
while gravity lives in 
a larger (10 or 11) dimensional bulk of the space-time. This ``Brane World'' 
picture\footnote{For recent developments, see, {\em e.g.}, 
\cite{witt,lyk,TeV,dienes,3gen,anto,BW}.
The brane 
world picture in the effective field theory context 
was discussed in \cite{early}. TeV-scale compactifications were originally
discussed in \cite{Ant} in the context of supersymmetry breaking.} 
{\em a priori} appears
to be a viable scenario, and, based on considerations of gauge and 
gravitational coupling unification, dilaton stabilization and weakness 
of the Standard Model gauge couplings, in \cite{BW} it was actually argued 
to be a likely description of nature. In particular, these phenomenological 
constraints seem to be embeddable in the brane world scenario 
(with the Standard Model fields living on branes with $3<p<9$), which therefore
might provide a coherent picture for describing our universe 
\cite{BW}. 
This is largely due to a much higher degree of flexibility of the brane 
world scenario compared with, say, the old perturbative heterotic framework. 

{}An important question arising in the brane world context is the issue of 
brane stabilization. In particular, in many four dimensional ${\cal N}=1$ 
supersymmetric Type I/Type I$^\prime$ vacua there are flat directions 
corresponding to the positions of branes either with respect to each other or
to the orientifold planes. Perturbatively, therefore, branes are not 
stabilized in such vacua. Upon supersymmetry breaking we might generically 
expect brane stabilization to occur. However, brane
stabilization can sometimes occur via supersymmetry preserving non-perturbative
dynamics. The purpose of this note is to illustrate some possible mechanisms of
brane stabilization in the brane world context. In particular, brane 
stabilization in the mechanisms we discuss here is due to non-perturbative
gauge dynamics in the world-volume of the branes, and can be supersymmetry
preserving as well as supersymmetry breaking dynamics. In the latter case
the VEV of the field whose F-term breaks 
supersymmetry is precisely the inter-brane distance or the distance between
the branes and an orientifold plane. For illustrative purposes we present 
{\em explicit} string models where these mechanisms are realized. These 
examples, which are consistent string vacua (with four non-compact and six
compact space-time dimensions) satisfying the requirements of
tadpole and anomaly cancellation, 
were originally constructed in \cite{class}, and here we briefly
review these constructions. In particular, the models we consider in this 
paper are four dimensional ${\cal N}=1$ supersymmetric orientifolds 
with non-trivial NS-NS $B$-flux.
 
{}Here we should emphasize that in the following discussions we will 
assume that
dilaton is somehow stabilized via some other dynamics whose relevant scale is
higher than the strong scale of the gauge dynamics responsible for brane
stabilization. We can then safely integrate dilaton out and consider
brane stabilization for a fixed gauge coupling determined by the dilaton VEV.
On the other hand,
if dilaton stabilization is not achieved, one runs into the 
usual runaway problem - the ground state of the theory is located at infinitely
weak coupling where branes are destabilized and supersymmetry is restored
(in the cases where brane stabilization is accompanied by supersymmetry 
breaking). The examples we discuss in this note, therefore, should only be
understood as illustrative (toy) representatives of ${\cal N}=1$ gauge 
theories (arising from consistent string vacua) where non-perturbative 
dynamics can stabilize the VEV of a field 
which in the brane world context is interpreted as measuring the
separation between branes or branes and orientifold planes.      

{}Let us begin our discussion with the case where brane stabilization is
accompanied by supersymmetry breaking. Thus, consider an ${\cal N}=1$
supersymmetric gauge theory with $SU(4)$ gauge group and matter consisting of
one chiral superfield $\Phi$ transforming in the two-index antisymmetric
irrep ${\bf 6}$ of $SU(4)$. Note that this theory is anomaly free. There is
a flat direction in this theory along which $\Phi$ acquires a VEV as follows:
$\langle\Phi\rangle=
\mbox{diag}(X\epsilon,X\epsilon)$, where $\epsilon\equiv i\sigma_2$ is
a $2\times 2$ antisymmetric matrix ($\sigma_2$ is a Pauli matrix). The gauge
group along this flat direction is broken down to $Sp(4)\approx SO(5)$, and the
only matter left is a singlet of $Sp(4)$ (which is precisely the chiral 
superfield acquiring the VEV $X$). Thus, below the mass scale $X$ we have pure
${\cal N}=1$ supersymmetric gauge theory with the gauge group $Sp(4)$ plus 
the singlet.

{}Supersymmetry in this model is actually broken. In fact, this is already
the case in the context of global supersymmetry.
The reason for this is that the gaugino condensate in the $Sp(4)$ gauge theory
depends non-trivially on $X$. This can be seen as follows. The gauge coupling
running starts at the string scale $M_s$ (which we will identify as the UV
cut-off of the theory). Between the scales $M_s$ and $X$ (where we assume
$X\ll M_s$) the gauge coupling of the $SU(4)$ gauge theory 
runs with the one-loop $\beta$-function 
coefficient $b_0=11$. Below the mass scale $X$ the gauge coupling of the
$Sp(4)$ gauge theory runs with the one-loop $\beta$-function coefficient
${\widetilde b}_0=9$. Matching the two gauge couplings at the scale $X$ then
gives the following strong scale ${\widetilde \Lambda}$ of the $Sp(4)$ theory: 
\begin{equation}
 {\widetilde \Lambda}=\Lambda^{b_0/{\widetilde b}_0}
 X^{1-b_0/{\widetilde b}_0} = \Lambda^{11/9}X^{-2/9}~,
\end{equation}
where $\Lambda\equiv M_s\exp(-2\pi S/b_0)$ is the strong scale of
the original $SU(4)$ gauge theory. (Here $S\equiv 4\pi/g^2+i\theta/2\pi$ is the
dilaton VEV which determines the $SU(4)$ gauge coupling $g$ at the string scale
as well as the vacuum angle $\theta$.) The gaugino condensation in the $Sp(4)$
theory generates the following superpotential\footnote{Note that for a pure
${\cal N}=1$ supersymmetric gauge theory with a {\em simple} 
gauge group $G$ the 
non-perturbative superpotential generated by the gaugino condensation 
(in the ${\overline {\mbox{DR}}}$ subtraction scheme) is
given by $W=h_G \Lambda_G^3$, where $h_G$ is the dual Coxeter number of
the group $G$ ($h_G=N$ for $G=SU(N)$, $h_G=N-2$ for $G=SO(N)$, $h_G=
N+1$ for $G=Sp(2N)$, and in our conventions $Sp(2N)$ has rank $N$), and 
$\Lambda_G\equiv \exp(-2\pi S/b_0(G))$ is the strong scale of the theory
(here $b_0(G)=3h_G$).}
\begin{equation}\label{sup}
 W=3{\widetilde \Lambda}^3=3\Lambda^{11/3} X^{-2/3}~.
\end{equation}
Note that this superpotential has runaway behavior in $X$ so that for any 
finite $X$ supersymmetry is broken. In the context of global supersymmetry
the vacuum is located at infinite $X$ where supersymmetry is restored. In the
locally supersymmetric context, however, the situation is quite different.

{}The key point here is the following. Thus, as was pointed out in 
\cite{runaway}, once such a theory with a runaway direction is coupled to
supergravity, generically there is a natural shut-down scale for 
runaway directions, namely, the Planck scale. More precisely, this is
the case for a certain large class of K{\"a}hler potentials, which, in
particular, includes K{\"a}hler potentials of minimal type. Local 
supersymmetry breaking and stabilization of the runaway directions
is then due to the interplay between non-perturbatively generated 
runaway superpotential and tree-level K{\"a}hler potential.

{}Let us illustrate the above point by considering a theory in which there
is a runaway superpotential $W=c/X^\alpha$ generated via some
non-perturbative dynamics. Let us assume that the K{\"a}hler potential for $X$
is minimal: $K=XX^*$. For simplicity in the following we will adapt the
units where the reduced Planck scale $M_P=1/\sqrt{8\pi G_N}$
is set to one, so that the scalar 
potential reads:
\begin{equation}
 V=e^K\left(K^{-1}_{XX^*}|F_X|^2 -3 |W|^2\right)~, 
\end{equation}  
where the F-term $F_X=W_X+K_X W$. Let $\rho\equiv XX^*$. Then we have
\begin{equation}
 V(\rho)=|c|^2 e^\rho\left({\alpha^2\over \rho^{\alpha+1}}-
 {2\alpha+3\over\rho^\alpha}+{1\over \rho^{\alpha-1}}\right)~.
\end{equation}
The extrema of this scalar potential are located at $\rho_0=\alpha$ and
$\rho_\pm=\alpha+1\pm\sqrt{\alpha+1}$. Here we are assuming that $\alpha>0$.
Then the extremum at $\rho=\rho_0$ corresponds to a {\em supersymmetric}
maximum with vanishing F-term. The extrema at $\rho=\rho_\pm$ correspond to
minima with broken supersymmetry. Thus, in such a theory $X$ is indeed
stabilized and supersymmetry is broken. Note that in the runaway superpotential
in (\ref{sup}) generated in the $SU(4)$ gauge theory with one chiral 
${\bf 6}$ which we discussed above we have $\alpha=2/3$. 

{}Note, however, that the superpotential (\ref{sup}) is only valid for
$X\ll M_s$, whereas the above minima correspond to the values of $X\sim
M_P$ (here we can assume for simplicity that the reduced Planck scale and
the string scale are of the same order of magnitude). So, strictly 
speaking, the above quantitative analysis is not valid. However, the 
qualitative conclusion that $X$ is stabilized and supersymmetry is broken is
still correct. The reason for this is the following. Note that
for $X\gg M_s$ we would have pure $Sp(4)$ super-Yang-Mills theory below the 
cut-off scale $M_s$. Then the superpotential generated in this case would
be independent of $X$: $W=3{\widetilde \Lambda}^3$, where
${\widetilde \Lambda}\equiv 
M_s \exp(-2\pi S/{\widetilde b}_0)$. The corresponding scalar
potential (in the assumption of the minimal K{\"a}hler potential) is then
given by\footnote{Here we should point out that for $X\gg M_s$ we must also
take into account the fact that generically dilaton may not be constant,
but vary logarithmically in the two transverse directions corresponding to
the motion of the branes. This would modify the $X$ dependence in $V$, but
it is not difficult to see that the qualitative conclusions on brane
stabilization at $X\sim M_s$ would still remain correct.}   
\begin{equation}
 V(\rho)=|W|^2e^\rho(\rho-3)~.
\end{equation}
This scalar potential can be seen to stabilize $\rho$ at $\rho=2$, where
the F-term is non-zero, hence supersymmetry breaking. For intermediate
values of $X$ (that is, $X\sim M_s$), the superpotential is a non-trivial
function of $X$ interpolating between $\propto X^{-2/3}$ for small $X$ 
and a constant for large $X$. (This interpolating function depends on the
details of various {\em threshold} corrections around the 
scale $M_s$.) From the limiting behavior, however, it is clear 
that $X$ is stabilized and supersymmetry is broken (subject to the
appropriate assumptions on the K{\"a}hler potential). As we have already 
mentioned, in the following we will present an explicit string model whose
low energy effective field theory contains the $SU(4)$ gauge theory with one
chiral ${\bf 6}$ discussed above, and where the VEV $X$ is identified with the 
separation between D-branes and the corresponding orientifold plane.

{}Next, we would like to discuss an example where brane stabilization occurs
via supersymmetry preserving dynamics. Thus, consider an ${\cal N}=1$ 
supersymmetric gauge theory with $SU(4)\otimes SU(4)$ gauge group and matter
consisting of one chiral superfield $\Phi$ transforming in 
$({\bf 10},{\bf 1})$ of $SU(4)\otimes SU(4)$, one chiral superfield 
$Q$ transforming in $({\overline {\bf 4}},{\bf 4})$, and one chiral superfield
${\widetilde Q}$ transforming in $({\overline {\bf 4}},{\overline{\bf 4}})$.
Note that this theory is anomaly free.
We will assume that there is a tree-level superpotential given by
\begin{equation}\label{tree}
 W_{tree}=\lambda \Phi Q{\widetilde Q}~,
\end{equation} 
where $\lambda$ is the corresponding Yukawa coupling. There are three flat
directions in this model. One corresponds to the field $Q$ acquiring a non-zero
VEV. The gauge group along the flat direction $\langle Q\rangle\not=0$ is
broken down to $SU(4)_{diag}\subset SU(4)\otimes SU(4)$. After the gauge 
symmetry breaking we have a singlet of $SU(4)_{diag}$ coming from $Q$, and 
also one chiral ${\bf 6}$ of $SU(4)_{diag}$ coming from ${\widetilde Q}$. The
rest of the fields in $Q$ are eaten in the super-Higgs mechanism, whereas
the rest of the fields in ${\widetilde Q}$ pair up with those in $\Phi$ and
become heavy via the corresponding coupling in the tree-level 
superpotential (\ref{tree}). Thus, along this flat direction we have the 
$SU(4)$ gauge theory with one chiral ${\bf 6}$ plus a singlet. This is the 
same theory as that already discussed above except for the presence of the 
extra singlet. One can analyze this case in a similar fashion, however, and
it is not difficult to see that VEV stabilization is accompanied by 
supersymmetry breaking just as in the previous case. Note that the second
flat direction along which $\langle {\widetilde Q}\rangle\not=0$ gives 
rise to the same theory as above. 

{}In the following we will therefore be interested in the third flat direction 
along which $\langle \Phi\rangle={\mbox{diag}}(X,X,X,X)$ and the gauge group is
broken down to $SO(4)\otimes SU(4)$. The only matter left is a singlet
(coming from $\Phi$) which is precisely the chiral superfield acquiring the
VEV $X$. The rest of the fields in $\Phi$ are eaten in the super-Higgs 
mechanism, and the fields $Q,{\widetilde Q}$ acquire mass $\lambda X$ 
due to the corresponding coupling in the tree-level superpotential 
(\ref{tree}). Thus, at low energies we have gaugino condensates in both $SO(4)
\approx SU(2)\otimes SU(2)$ and $SU(4)$ subgroups. Matching the gauge couplings
of $[SU(2)\otimes SU(2)]\otimes SU(4)$ with 
the original gauge couplings of $SU(4)\otimes
SU(4)$ at the corresponding threshold scales, 
we obtain the following non-perturbative 
superpotential on this branch of the classical moduli space
(here we assume that gauge couplings of both $SU(4)$ subgroups are the same at
the string scale):
\begin{equation}\label{sup1}
 W=2  \lambda^4 X^{-2}\Lambda^5 + 4 \lambda X(\Lambda^\prime)^2~,
\end{equation} 
where $\Lambda\equiv M_s \exp(-2\pi S/
b_0)$, and $\Lambda^\prime\equiv M_s \exp(-2\pi S/
b_0^\prime)$. Here $b_0=5$ and $b_0^\prime=8$ are the one-loop 
$\beta$-function coefficients of the first respectively second $SU(4)$ 
subgroups above the corresponding threshold scales.

{}In fact, we can generalize the above model as follows. Thus, consider an 
${\cal N}=1$ 
supersymmetric gauge theory with $SU(M)\otimes SU(N)$ gauge group and matter
consisting of one chiral superfield $\Phi$ transforming in 
$({\bf R}_\eta,{\bf 1})$ of $SU(M)\otimes SU(N)$, one chiral superfield 
$Q$ transforming in $({\overline {\bf M}},{\bf N})$, and one chiral superfield
${\widetilde Q}$ transforming in $({\overline {\bf M}},{\overline{\bf N}})$.
Here ${\bf R}_\eta={\bf S}$ for $\eta=+1$, and ${\bf R}_\eta={\bf A}$ 
for $\eta=-1$, where ${\bf S}$ is the two-index symmetric ($M(M+1)/2$ 
dimensional) representation of $SU(M)$, while ${\bf A}$ is the two-index
antisymmetric ($M(M-1)/2$ dimensional) representation of $SU(M)$. Anomaly 
cancellation requires that $M=2N-4\eta$. We will assume that there is a 
tree-level superpotential given by (\ref{tree}). Along the flat direction
$\langle \Phi\rangle \not=0$ the gauge group is broken down to
$G_\eta(M)\otimes SU(N)$, where $G_\eta(M)=SO(M)$ for $\eta=+1$, and   
$G_\eta(M)=Sp(M)$ for $\eta=-1$. The only matter left is a singlet coming from 
$\Phi$. The gaugino condensation in both $G_\eta(M)$ and $SU(N)$ subgroups
gives rise to the following non-perturbative superpotential:
\begin{equation}\label{non-pert}
 W=C(N,\eta) \lambda^{N\over N-3\eta}
 X^{-{N-2\eta\over N-3\eta}}\Lambda^{4N-11\eta\over N-3\eta}
 +N(\lambda X)^{2-{4\eta\over N}}(\Lambda^\prime)^{1+{4\eta\over N}}~,
\end{equation}
where $\Lambda^{4N-11\eta}\equiv M_s\exp(-2\pi S)$, and 
$(\Lambda^\prime)^{N+4\eta}\equiv M_s\exp(-2\pi S)$ (and we are assuming that
gauge couplings of both $SU(M)$ and $SU(N)$ are the same at the string scale).
Here $C(N,\eta)=N-3\eta$ except when $N=4,\eta=+1$. In the latter case 
$G_\eta(M)=SO(4)\approx SU(2)\otimes SU(2)$ (that is, $G_\eta(M)$ is not 
simple), and $C(N,\eta)=2$ (instead of 1) as we have to add the contributions
of gaugino condensates in both $SU(2)$ subgroups. (Here we are assuming
that the phases of the gaugino condensates are the same for both $SU(2)$
subgroups.) 
We note that in deriving the
superpotential (\ref{non-pert}) some care is needed. In particular, for 
$\eta=-1$ we have the breaking $SU(M)\supset Sp(M)$, and the $Sp(M)$ gauge 
coupling at the scale $X$ is the same as that of $SU(M)$. However, for
$\eta=+1$ we have the breaking $SU(M)\supset SO(M)$, and the $SO(M)$ gauge 
coupling ${\widetilde g}(X)$ at the scale $X$ is related to the $SU(M)$
gauge coupling $g(X)$ at the same scale via $2{\widetilde g}^2(X)=g^2(X)$.
This is due to the special embedding of $SO(M)$ into $SU(M)$. (Thus, in the
perturbative heterotic language $SO(M)$ in this case is realized via a level-2
current algebra.) In particular, this is important in obtaining the correct 
powers of $X$ and $\Lambda$ as well as the numerical coefficient $C(N,\eta)$
in the first term in (\ref{non-pert}).   
 
{}Let us go back to the non-perturbative superpotential (\ref{sup1}) in the
$SU(4)\otimes SU(4)$ example discussed above. Let us first consider the 
globally supersymmetric setup. The F-term of $X$ vanishes for
\begin{equation}\label{min}
 X=\lambda \left({\Lambda^5\over (\Lambda^\prime)^2}\right)^{1/3}=
 \lambda M_s \exp(-\pi S/2)~.
\end{equation}
For a large enough dilaton VEV $S$ (that is, for small enough gauge coupling
at the string scale\footnote{Note that large $S$ in our conventions does not
necessarily imply weak string coupling. Thus, as was pointed out in \cite{BW},
the string coupling can be of order 1, while the gauge coupling (at the string 
scale) of a gauge theory arising from a set of $p$-branes compactified on
a $p-3$ dimensional space can be small if the corresponding compactification
volume $V_{p-3}$ is large in the string units. This volume is absorbed in the
above definition of the dilaton VEV $S$.}) we have $X\ll M_s$, so that 
the above quantitative analysis is valid in the globally supersymmetric 
context. That is, the VEV $X$ is stabilized via supersymmetry {\em preserving}
dynamics. Here we would like to address the issue of whether the same is true
once we couple this system to supergravity.

{}To answer the above question, consider a generic system of superfields 
$\Phi_i$ with the superpotential $W$. Let $K(\Phi,{\overline \Phi})$ be the 
K{\"a}hler potential, which we will assume to be non-singular for non-zero 
finite values of $\Phi_i$. The scalar potential is given by (once again,
here we set the reduced Planck scale $M_P$ to one):
\begin{equation}\label{GenScalar}
 V=e^K\left(G_{i{\bar j}} F_i (F_j)^*-3WW^*\right)~,
\end{equation}
where the summation over repeated indices is understood, 
$G_{i{\bar j}}\equiv K^{-1}_{i{\bar j}}$ is the inverse of the 
K{\"a}hler metric, $F_i\equiv W_i+K_iW$ are the F-terms for the 
fields $\Phi_i$, and $W_i\equiv\partial W/\partial \Phi_i$.

{}Note that if all of the F-terms $F_i$ are vanishing, local supersymmetry 
is unbroken. It is easy to verify that a point with $F_i\equiv0$ corresponds 
to an extremum of the scalar potential. (In this case $V_i=V_{\bar i}=0$.) 
We would like to investigate the conditions under which this extremum is 
actually a local minimum of the scalar potential. To do this we need the 
following quantities evaluated at the point $F_i\equiv0$:
\begin{eqnarray}
 V_{ij}&=&-e^K F_{ij}W^*~,\\
 V_{i{\bar j}}&=&e^K \left[G_{i^\prime{\bar j}^\prime} F_{i^\prime i}
 (F_{j^\prime j})^*-2K_{i{\bar j}}
 WW^*\right]~.
\end{eqnarray}
Here $F_{ij}\equiv(F_i)_j=W_{ij}+(K_{ij}-K_i K_j) W$. Note that if, say, 
the absolute values of all the eigenvalues of the matrix $W_{ij}$ are much 
larger\footnote{One can easily write down the precise condition for the 
minimum which does not require the absolute values of the eigenvalues of 
the matrix $W_{ij}$ be much larger than $\vert W\vert$, but satisfy certain 
inequalities.} than $\vert W\vert$ at the point $F_i\equiv0$, then this point 
corresponds to a local minimum of the scalar potential\footnote{Here we 
assume that the K{\"a}hler potential and its derivatives are 
of order 1 or smaller at
$\Phi_i=\Phi^{(0)}_i$, which is usually the case.}. 

{}The above discussion has the following implications for the fate of globally 
supersymmetric minima corresponding to the superpotential $W$ in the local 
supergravity context. Suppose the 
F-flatness conditions $W_i=0$ in the global setting have a solution with 
all the VEVs stabilized at $\Phi_i=\Phi^{(0)}_i$. Furthermore, suppose 
that the eigenvalues of the matrix $W_{ij}(\Phi^{(0)})$ are all non-zero, and 
their absolute values are much larger than $\vert W(\Phi^{(0)})\vert$. Then 
in the vicinity of the point $\Phi^{(0)}_i$ there is another point 
$\Phi^{(1)}_i$ which is a locally
supersymmetric minimum of the scalar potential (\ref{GenScalar}).
The proof of this statement is very simple. First, the F-flatness 
conditions in the local setting are given by $W_i(\Phi)=-K_i(\Phi,
{\overline \Phi})W(\Phi)$. Since $W_{ij}(\Phi^{(0)})$ is non-singular, 
these equations have a solution at $\Phi_i=\Phi^{(1)}_i$, where 
$\Phi^{(1)}_i=\Phi^{(0)}_i-M_{ij}(\Phi^{(0)})K_j
(\Phi^{(0)},\overline{\Phi}^{(0)}) W(\Phi^{(0)})+
{\cal O}(\epsilon^2)$, and $\epsilon<<1$ is the absolute value
of the largest eigenvalue 
of the matrix $M_{ij}(\Phi^{(0)}) W(\Phi^{(0)})$, where $M_{ij}
(\Phi^{(0)})$ is the matrix inverse to $W_{ij}(\Phi^{(0)})$. Moreover, 
according to our discussion above, the point $\Phi^{(1)}_i$ corresponds 
to a local minimum. Here we note that the physical reason for this 
is clear - since in the globally supersymmetric setup all the fields
are heavy at the minimum, there is no Goldstino candidate required for
local supersymmetry breaking once we couple the system to supergravity.

{}Let us now go back to the superpotential (\ref{sup1}), and see whether
the aforementioned conditions are satisfied at the global minimum given by
(\ref{min}). Thus, we have
\begin{equation}
 |W_{XX}/W|=2|\lambda|^{-2}M_s^{-2}\exp(\pi {\mbox{Re}}(S))\gg 1~.
\end{equation}
Here we are using the $M_P=1$ units, and we are assuming that the Yukawa
coupling is of order of the gauge coupling at the string scale: 
$|\lambda|\sim g$. Thus, in this model the VEV $X$ is indeed stabilized 
without supersymmetry breaking, and the stabilized value of $X$ is much
smaller than $M_s$ (for large enough values of $S$).

{}Next, as we promised in the beginning, we are going to give the 
explicit string construction of the above models. In fact, these models
were already constructed in \cite{class} where a more detailed discussion 
can be found. Here we will briefly review the construction in \cite{class}, 
and identify various flat directions in the classical moduli space with
the motion of D-branes in the corresponding transverse directions.

{}Consider Type I compactified on the following orbifold Calabi-Yau
three-fold with $SU(3)$ holonomy: ${\cal M}_3\equiv (T^2\otimes T^2\otimes
T^2)/({\bf Z}_3\otimes {\bf Z}_3)$. Let $g_1$ and $g_2$ be the generators
of the two ${\bf Z}_3$ subgroups. Their action on the complex coordinates
$z_1,z_2,z_3$ parametrizing the three two-tori is given by:
\begin{eqnarray} 
 &&g_1 z_1=\omega z_1~,~~~g_1 z_2=\omega^{-1} z_2~,~~~g_1 z_3=z_3~,\\
 &&g_2 z_1=z_1~,~~~g_2 z_2=\omega  z_2~,~~~g_2 z_3=\omega^{-1} z_3~.
\end{eqnarray}
Here $\omega\equiv\exp(2\pi i/3)$. This Calabi-Yau three-fold has the Hodge
numbers $(h^{1,1},h^{2,1})=(84,0)$, so there are 84 chiral supermultiplets
in the closed string sector. In the open string sector we have D9-branes
only whose number depends on the rank $b$ of the NS-NS $B$-field $B_{ij}$
($i,j=1,\dots,6$ label the real compact directions corresponding to
complex coordinates $z_1,z_2,z_3$). Note that since $B_{ij}$ is
antisymmetric, $b$ can only take even values $0,2,4,6$. The untwisted
tadpole cancellation conditions then imply that we have $2^{5-b/2}$ 
D9-branes \cite{bij}. In the following we are going to be interested in 
the cases $b=2$ and $b=4$.

{}${\bullet}$ $b=2$. We have 16 D9-branes. 
The orientifold projection is of the $Sp$ type.
The solution to the twisted tadpole cancellation conditions 
(up to equivalent representations) is given by\footnote{Here we work with
$2^{4-b/2}\times 2^{4-b/2}$ (rather than $2^{5-b/2}\times 2^{5-b/2}$)
Chan-Paton matrices as we choose not to count the orientifold images of the
corresponding D-branes.} 
\cite{class}:
\begin{eqnarray}
 &&\gamma_{g_1,9}={\mbox{diag}}(\omega {\bf I}_4,\omega^{-1} {\bf I}_4)~,\\
 &&\gamma_{g_2,9}={\mbox{diag}}(\omega {\bf I}_2,{\bf I}_2,\omega^{-1} 
 {\bf I}_2,{\bf I}_2)~.
\end{eqnarray}
Here we have chosen $B_{12}=1/2$, $B_{34}=B_{56}=0$.
The gauge group is $U(4)\otimes U(4)$. 
The matter consists of the following chiral superfields: $\Phi=({\bf 10},
{\bf 1})(+2,0)$, $Q=({\overline {\bf 4}},{\bf 4})(-1,+1)$, ${\widetilde Q}=
({\overline {\bf 4}},{\overline {\bf 4}})(-1,-1)$, where the $U(1)$ charges
are given in parentheses.
The tree-level superpotential, which can be derived using the standard 
techniques, reads:
\begin{equation}\label{tree1}
 {\cal W}=\lambda \Phi Q{\widetilde Q}~.
\end{equation}
Note that this is precisely the second model discussed above except for the
extra $U(1)$'s. In fact, the first $U(1)$ is anomalous. It is broken by
VEVs of some of the twisted closed string sector
chiral multiplets which transform 
non-linearly under the anomalous $U(1)$.
The second $U(1)$ is anomaly free,
and it is not difficult to see that it has no effect on the above discussion of
VEV stabilization.

{}Note that in the above construction we have D9-branes. We can T-dualize
to arrive at the corresponding Type I$^\prime$ vacuum with, say,
D3-branes. Then the space transverse to D3-branes is ${\cal M}_3$. The 
point in the moduli space with the unbroken gauge group corresponds to the
brane configuration where all D3-branes are sitting on top of the
corresponding orientifold plane. The flat directions along which $\langle
\Phi\rangle\not=0$, $\langle {\widetilde Q}\rangle\not=0$ and
$\langle Q\rangle\not=0$ then correspond to the branes moving off the
orientifold plane in the $z_1$, $z_2$ and $z_3$ directions,
respectively. Thus, if $\Phi\not=0$ then motion in the $z_2,z_3$ directions
is not allowed due to the corresponding F-flatness conditions implied by 
the superpotential (\ref{tree1}). On the other hand, as we have discussed
above, the VEV of $\Phi$ is stabilized via supersymmetry preserving
non-perturbative dynamics on this branch of the classical moduli space, so
that D3-branes are stabilized at a finite distance from the orientifold
plane.

${\bullet}$ $b=4$. We have 8 D9-branes. The orientifold projection 
is of the $SO$ type.
The solution to the twisted tadpole cancellation conditions is given by
\cite{class} :
\begin{eqnarray}
 &&\gamma_{g_1,9}={\bf I}_4~,\\
 &&\gamma_{g_2,9}={\mbox{diag}}(\omega {\bf I}_2,\omega^{-1} {\bf I}_2)~.
\end{eqnarray} 
Here we have chosen $B_{12}=B_{34}=1/2$, $B_{56}=0$.
The gauge group is
$U(4)$. The matter consists of one chiral superfield $\Phi={\bf
6}(+2)$. Note that we have anomalous $U(1)$ in this model which is broken
by VEVs of the corresponding twisted closed string sector chiral 
multiplets\footnote{Note that in this model the corrections to the 
K\"ahler potential due to the anomalous $U(1)$ breaking are important 
for the quantitative analysis of brane stabilization. They, however, do not
affect the qualitative picture of brane stabilization.}.
Note that there are no renormalizable couplings in this case. This model is
then the same as the first model we discussed above. Once we T-dualize so
that we have D3- instead of D9-branes, the motion of D3-branes 
in the $z_3$ direction is described by the $\Phi$ VEV. Note that D3-branes
are frozen in the $z_1,z_2$ directions as the corresponding moduli are
absent. As we saw above, the VEV of $\Phi$ in this model is stabilized via
supersymmetry breaking non-perturbative dynamics, so that D3-branes are
stabilized at a finite distance from the orientifold plane.
 
{}Here we should point out that the above open string spectra correspond to
perturbative (from the orientifold viewpoint) sectors. However, unlike in
some other cases (examples of which were recently discussed in \cite{np}), 
non-perturbative orientifold sector states in these models become heavy and 
decouple once we blow up the orbifold singularities (which is required to
break anomalous $U(1)$'s) along the lines of \cite{blowup}.

{}Before we finish this note, we would like to make a few comments. First,
it would be interesting to see whether in models of this type 
log-like corrections to the
K{\"a}hler potential could stabilize branes at distances {\em large}
compared with the string length along the lines of \cite{dvali}. Also, the
above discussion assumed dilaton stabilization. In a more realistic setup
dilaton would have to be stabilized via the standard ``race-track''
mechanism \cite{kras}, or via a quantum modification of the moduli space
\cite{gia} (also see \cite{dudas}), 
or else via some other mechanism. It would be interesting to
understand these issues more explicitly in the brane world context (for a
general discussion see \cite{BW}). 

{}Finally, we should point out that brane stabilization mechanisms similar
to those discussed above were considered
in \cite{ooguri} in the {\em non-compact} context
where gravity decouples from the gauge theory dynamics. Here we consider
consistent {\em compact} tadpole-free orientifold models. Brane
stabilization in the latter context was also discussed in
\cite{lykken}, where branes were argued to be stabilized via an interplay
between non-perturbatively generated superpotential and an anomalous
$U(1)$. However, as we have mentioned above, anomalous $U(1)$'s in the
orientifold vacua are generically broken by VEVs of the corresponding
twisted closed string sector chiral multiplets, which ameliorates the
effect of the anomalous $U(1)$'s leaving behind the usual non-perturbative 
runaway superpotential (as in the first model discussed in this note).   

{}I would like to thank Gia Dvali and Gregory Gabadadze for
discussions. This work was supported in part by the grant NSF PHY-96-02074, 
and the DOE 1994 OJI award. I 
would also like to thank Albert and Ribena Yu for 
financial support.

\end{document}